\documentclass[aps,prl,reprint,amsmath,amssymb,amsfonts,floatfix, showpacs,intlimits, euspript,superscriptaddress  ]{revtex4-1}
\usepackage{graphicx}
\usepackage{textcomp}
\usepackage[caption=false]{subfig}
\usepackage{xcolor}
\usepackage{siunitx}
\usepackage{esint}
\usepackage{accents}
\usepackage{bbm}
\usepackage{mathtools}
\usepackage{tikz}
\usepackage{tikz-3dplot}
\usepackage{layouts}

\usetikzlibrary{calc,3d}

\renewcommand{\vec}[1]{{\boldsymbol{#1}}}

\newcommand{\cross}{\times}

\newcommand{\dsv}[1]{{\vec{\mathbbm{#1}}}}

\newcommand{\z}{\dsv{z}}
\newcommand{\y}{\dsv{y}}
\newcommand{\x}{\dsv{x}}

\newcommand{\Ms}{M_s}

\newcommand{\vl}{\vec{l}}

\newcommand{\W}{\mathcal{W}}
\newcommand{\vS}{\vec{S}}

\DeclareMathOperator{\sign}{sign}

\DeclareMathOperator{\Real}{Re}

\newcommand{\vs}{\vec{s}}

\newcommand{\wa}{\omega_a}

\newcommand{\jsp}{\vec{j}_\text{sp}}
\newcommand{\jsh}{\vec{j}_\text{sh}}

\newcommand{\wex}{\omega_\text{ex}}

\newcommand{\tSH}{\theta_\text{SH}}
\newcommand{\aG}{\alpha_\text{G}}
\newcommand{\electron}{\si{\elementarycharge}}
\newcommand{\wafmr}{\omega_\text{afmr}}

\definecolor{amber}{rgb}{1.0, 0.75, 0.0}
\definecolor{current}{rgb}{0.64, 0.0, 0.0}
\definecolor{spincurrent}{rgb}{0.03, 0.27, 0.49}
\definecolor{stt}{rgb}{0.55, 0.0, 0.55}
\definecolor{Neel}{rgb}{0.09, 0.45, 0.27}

\begin{document}

\title{Sub-terahertz ferrimagnetic spin-transfer torque oscillator}

\date{\today}

\author{Ivan Lisenkov}
\email[]{ivan@lisenkov.com}
\affiliation{Winchester Technologies LLC, Burlington, MA}

\author{Roman~Khymyn}
\affiliation{Department of Physics, University of Gothenburg, 41296 Gothenburg, Sweden}
\affiliation{NanOsc AB, 16440 Kista, Sweden}
\author{Johan \r{AA}kerman}
\affiliation{Department of Physics, University of Gothenburg, 41296 Gothenburg, Sweden}
\affiliation{NanOsc AB, 16440 Kista, Sweden}

\author{Nian X. Sun}
\affiliation{Electrical and Computer Engineering Department, Northeastern University, Boston, MA}

\author{Boris A. Ivanov}
\affiliation{Institute of Magnetism, NASU and MESU, Kiev, 03142, Ukraine}
% \affiliation{Institute of Magnetism, NASU and MESU, Kiev, 03142, Ukraine}
\affiliation{National Univeristy of Science and Technology, ``MISiS'', Moscow, Russian Federation}

\begin{abstract}
  A theory of magnetization dynamics in ferrimagnetic materials with antiparallel aligned spin sub-lattices under action of spin-transfer torques (STT) is developed. We consider magnetization dynamics in GdFeCo layers in two cases of magnetic anisotropy: easy plane and easy axis. We demonstrate that, (i) for the easy plane anisotropy the precession of the N\'{e}el vector is conical and the cone angle depends on the STT strength and the value of spin non-compensation, while the frequency of precession can reach sub-THz frequencies; (ii) for the easy axis anisotropy two regimes are possible: deterministic switching of the net magnetization and a conical sub-THz precession depending on the STT strength. 
\end{abstract}

\maketitle

Spin-transfer torques~\cite{bib:Slonczewski:1996} (STT) are widely used to control and excite magnetization dynamics in ferromagnetic materials (FM). Under certain conditions, the STT can overcome damping and set the magnetization in a steady-state precession~\cite{bib:Kiselev:2003,bib:Demidov:2012, bib:Slavin:2009}. The frequency of magnetization precession in FM is typically defined by the external magnetic field and experimentally achievable values usually lie in the GHz range~\cite{Chen2016procieee, bonetti2009spin}.

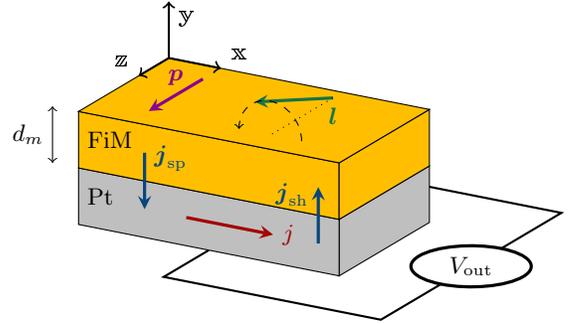
\begin{figure}

  \tdplotsetmaincoords{70}{120}
  \tdplotsetrotatedcoords{0}{90}{90}
    \begin{tikzpicture}[tdplot_main_coords,scale=0.8]
   
    \def\h{1.0}
    \def\w{5.0}
    \def\d{3.0}
    \def\offset{.5}
    \def\curroffset{0.4}
    \def\vradius{1.0}

    % Voltmeter
    \draw[thick] (0,-\w/2,0)--(\d/2,-\w/2, 0)--(\d/2,\w/3,0) -- (-3*\d/2, \w/3) -- (-3*\d/2,-\w/2, 0);

    \begin{scope}[canvas is yx plane at z=0]
      \draw[fill=white, very thick, draw=black] (\w/3, -\d/2) circle (\vradius) node[sloped]{$V_\text{out}$};
    \end{scope}

    % Pt layer
    \draw[fill=lightgray] (0,0,0)--(-\d,0,0)--(-\d,0,\h)--(0,0,\h)--cycle;
    \draw[fill=lightgray] (0,0,0)--(0,-\w,0)--(0,-\w,\h)--(0,0,\h)--cycle;
    \draw (0,-\w, \h/2) node[right]{Pt};

    % FiM layer
    \draw[fill=amber] (0,0,\h)--(0,-\w,\h)--(0,-\w,2*\h)--(0,0,2*\h)--cycle;
    \draw[fill=amber] (0,0,\h)--(-\d,0,\h)--(-\d,0,2*\h)--(0,0,2*\h)--cycle;
    \draw[fill=amber] (-\d,0,2*\h)--(0,0,2*\h) -- (0,-\w,2*\h) -- (-\d,-\w,2*\h) -- cycle;
    \draw (0,-\w, 3*\h/2) node[right]{FiM};

    % dimensions
    % \draw[<->] (0,-\w - \offset,0)-- node[midway, left]{$d_{\text{Pt}}$}(0,-\w -\offset,\h);
    \draw[<->] (0,-\w - \offset,\h)-- node[midway, left]{$d_{m}$}(0,-\w -\offset,2*\h);

    % current
    \draw[-stealth, very thick, current] (0, -\w*2/3 + \curroffset, \h/2) -- (0, -\w*1/3+\curroffset, \h/2) node[right] {$j$} ;

    % spin-pumping
    \draw[-stealth, very thick, spincurrent] (0, -\w*2/3 - \curroffset, 3*\h/2) -- node[anchor=south west] {$\jsp$}(0, -\w*2/3-\curroffset, \h/2)  ;

    % spin-hall
    \draw[stealth-, very thick, spincurrent] (0, - \curroffset, 3*\h/2) -- node[anchor=south east] {$\jsh$}(0, -\curroffset, \h/2)  ;

    % STT
    \draw[-stealth, very thick, stt] (-\d*4/5, -\w*4/5, 2*\h) -- node[above] {$\vec{p}$}(-\d/5, -\w*4/5, 2*\h);
    
    % Coordinate system
    \coordinate (Shift) at (-\d, -\w, 2*\h);
    \tdplotsetrotatedcoordsorigin{(Shift)}
    \draw[tdplot_rotated_coords, thick,->] (0,0,0) -- (1.0,0,0) node[anchor=south west]{$\x$};
    \draw[tdplot_rotated_coords, thick,->] (0,0,0) -- (0,1.0,0) node[anchor=north west]{$\y$};
    \draw[tdplot_rotated_coords, thick,->] (0,0,0) -- (0,0,1.0) node[anchor=south east]{$\z$};

    % N\'{e}el vector

    \coordinate (A) at (-\d/5, -\w/3, 2*\h);
    \tdplotsetrotatedcoordsorigin{(A)}

    \draw[tdplot_rotated_coords, -stealth, very thick, Neel] (0, 0, -\w*5/12)  
      node[below]{$\vl$} -- (-\d/5/2, \d/5*0.886, 0) ;
    \draw[tdplot_rotated_coords, -, dotted] (0, 0, -\w*5/12) -- (  0, 0, 0);

    \tdplotdrawarc[tdplot_rotated_coords, dashed, ->]{(0,0,0)}{\d/5}{0}{180}{anchor=north}{};

\end{tikzpicture}
\caption{Sketch showing the geometry of a ferrimagnetic spin-Hall oscillator}
\label{fig:geom}
\end{figure}

The action of STT is not limited to ferromagnetic materials, STT can act on any other magnetically ordered materials (xM): antiferromagnets (AFM)~\cite{bib:Jungwirth:2018, bib:Chen:2018,bib:Okuno:2019} and ferrimagnets (FiM)~\cite{bib:Roschewsky:2016}. It has been theoretically predicted, that magnetization sub-lattices of AFM materials under application of SST can precess with THz frequencies~\cite{bib:Gomonay:2010, bib:Gomonay:2014, bib:Gomonay:2017, bib:Khymyn:2017, bib:Cheng:2016}. A typical configuration of an AFM or FiM THz oscillator includes a bi-layer of a heavy metal (typically Pt) and an xM layer, see Fig.~\ref{fig:geom}. A current running through the metal generates a spin-current density $\jsh$ in the Pt layer via the spin-Hall effect. In its turn, the spin-current penetrates into the xM layer and induces STT, $\tau \vec{p}$, on the spin sub-lattices of xM. Vector $\vec{p}$, denoting the STT polarization, lies in the interface plane and perpendicular to the charge current. In the case of an uniaxial (easy axis or easy plane) AFM the application of STT forces the N\'{e}el vector, $\vl$, to rotate in a plane perpendicular to the STT polarization with a constant angular velocity~\cite{bib:Gomonay:2010, bib:Gomonay:2014, bib:Gomonay:2017, bib:Khymyn:2017, bib:Cheng:2016}:
\begin{equation}
   \dot\phi = \omega_r = \sigma j/\aG,
   \label{eq:omega_r}
\end{equation}
where $\sigma = \tSH/(2\electron S_T  d_m)$ is the STT ``efficiency'', $j$ is the charge current density in the Pt layer, $\electron$ is the elementary charge, $S_T$ is the total volume spin density of all sublattices, $d_m$ is the magnetic material thickness, $\tSH$ is the effective spin-Hall angle, and $\aG$ is the effective Gilbert damping constant~\cite{bib:Gomonay:2014, bib:Khymyn:2017,bib:Cheng:2014, bib:Cheng:2016, bib:Zhang:2015, bib:Tserkovnyak:2002}. Damping in thin layers is defined by the intrinsic damping and a spin-pumping process, \emph{i.e.}~loss of the angular momentum back to the Pt from the magnetic system~\cite{bib:Tserkovnyak:2002, bib:Zhang:2015}. 

\begin{figure}
  \tdplotsetmaincoords{70}{120}
  \begin{tikzpicture}[tdplot_main_coords]

    % \tikzset{>=stealth}
    \coordinate (O) at (0,0,0);

    \draw[thick,->] (0,0,0) -- (3.5,0,0) node[anchor=north east]{$\x$};
    \draw[thick,->] (0,0,0) -- (0,3.5,0) node[anchor=north west]{$\y$};
    \draw[thick,->] (0,0,0) -- (0,0,3.5) node[anchor=south]{$\z$};

    \draw[ultra thick, -stealth, stt] (0,0,0) -- (0,0,3) node[left]{$\vec{p}$};

    \pgfmathsetmacro{\ax}{3}\pgfmathsetmacro{\ay}{0}\pgfmathsetmacro{\az}{2.6};
    \pgfmathsetmacro{\bx}{-2}\pgfmathsetmacro{\by}{0}\pgfmathsetmacro{\bz}{1.0};

    \tdplotdrawarc[dashed]{(0,0,0)}{\ay}{0}{360}{anchor=north}{};
    \tdplotsetrotatedcoords{0}{0}{80}

    \draw[thick, -stealth, red, tdplot_rotated_coords] (0,0,0) -- (\ax,\ay,\az) node[anchor=west]{$\dfrac{\vec{S}_1}{S_1+S_2}$};
    \draw[thick, -stealth, blue, tdplot_rotated_coords] (0,0,0) -- (\bx,\by,\bz) node[anchor=east]{$\dfrac{\vec{S_2}}{S_1+S_2}$};
    \draw[thick, -stealth, black!60!green, very thick, tdplot_rotated_coords] (\bx,\by,\bz) -- node[below]{$\vl$} (\ax,\ay,\az) ;
    % \draw[]
    \draw[dashed,black!60!green, tdplot_rotated_coords] (\bx,\by,\bz) -- (\bx,\by,0)  -- (\ax,\ay,0) -- (\ax,\ay,\az);

    \tdplotdrawarc[->, thick]{(O)}{1.5}{0}{80}{anchor=north}{$\phi$}

    \coordinate (Shift) at (0,0,\az/2 + \bz/2);
    % \tdplotsetrotatedcoords{-20}{10}{0}
    \tdplotsetrotatedcoordsorigin{(Shift)}
    \tdplotsetrotatedcoords{0}{90}{80}

    % \tdplotsetrotatedthetaplanecoords{180}
    \tdplotdrawarc[tdplot_rotated_coords, ->, thick]{(0,0,0)}{0.5}{90+10}{10}{anchor=south}{$\theta$}

    \end{tikzpicture}
    \caption{Schematic representation of the N\'{e}el vector $\vl$ rotation in a ferrimagnetic with uncompensated spins $\vS_1$ and $\vS_2$ under spin transfer torque $\vec{p}$}

    \label{fig:spherical}
\end{figure}
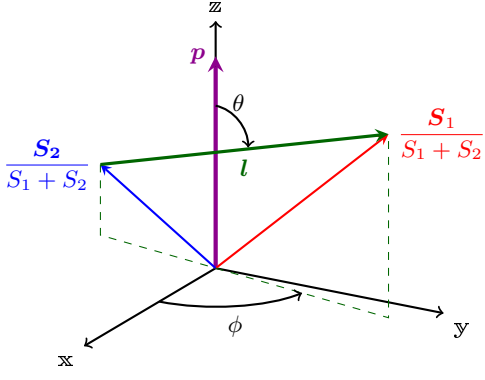

Since the magnetic moments in AFMs are compensated, extraction of the THz signal is difficult. One of the solutions is to use materials with weak ferromagnetism, which have a small spontaneous magnetic moment, rotation of which emits electromagnetic radiation~\cite{bib:Sulymenko:2017}. Another approach is to use the inverse spin-Hall effect (ISHE)~\cite{bib:Cheng:2014}. Indeed, a steady rotation of the N\'{e}el vector generates a spin-current $\jsp$ via the spin-pumping mechanism~\cite{bib:Tserkovnyak:2002}. The spin-current transfer through the interface between an AFM or FiM layer to a Pt layer can be written as:
\begin{multline}
  \jsp = \dfrac{\hbar g_r}{2\pi} \vl\cross\dfrac{d\vl}{dt} =\\
   \dfrac{\hbar g_r}{2\pi} \omega_r
   \left(\z\sin^2\theta - \sin\theta\cos\theta(\x+i\y)e^{-i\omega_rt}\right),
  \label{eq:jsp}
\end{multline}
where $\theta$ is the azimuthal angle of $\vl$, $g_r$ is the spin-mixing conductance on the interface, and the DC spin-current is polarized along the $\z$ axis (see Fig.~\ref{fig:geom} and~\ref{fig:spherical}). The AC components of the spin-current (and the AC component of the ISHE voltage) vanishes if $\vl$ rotates in a plane perpendicular to $\vec{p}$, \emph{i.e.}~if $\theta=\pi/2$. Unfortunately, this scenario, known as ``proliferation'', is realized in AFMs with uniaxial anisotropy, resulting in a zero AC output voltage~\cite{bib:Cheng:2016}. Several approaches have been proposed to overcome this difficulty, \emph{e.g.}~using materials with biaxial anisotropy~\cite{bib:Khymyn:2017, bib:Sulymenko:2018, bib:Khymyn:2018} or a nonlinear ISHE feedback~\cite{bib:Cheng:2016}.

In this Letter we demonstrate that in a FiM with nearly compensated spins, the spin dynamics is conical, \emph{i.e.}~$\theta\neq\pi/2$. The conical precession generates non-zero components of spin-current, in contrast with the case of a fully compensated AFM~\cite{bib:Khymyn:2017, bib:Cheng:2016}. Moreover, for an easy axis configuration, we demonstrate that depending of the value of the STT, the FiM sub-lattices may either switch their ground states or come into a steady conical precessional motion. As a model ferrimagnetic material we consider GdFeCo, which demonstrates low Gilbert damping and has a tunable configuration of magnetic anisotropy~\cite{bib:Ding:2013}. 

Magnetization dynamics in a ferrimagnetic material is governed by two coupled Landau-Lifshitz equations~\cite{bib:Ivanov:1983}:
\begin{equation}
  \hbar\dfrac{d\vS_i}{dt} = \left(\vS_i\cross\dfrac{\delta\W}{\delta\vS_i} + \dfrac{1}{S_i}\vS_i\cross \vS_i\cross\mathcal{R}_i\right),
  \label{eq:LL_coupled}
\end{equation}
where $i=1,2$ is the sub-lattice index, $\vS_i$ is the effective spin of each sub-lattice, $\W$ is the free energy per spin of the ferrimagnet, and $\mathcal{R}_i$ is the dissipative term. For uniaxial anisotropy the energy can be written as: $\W = \dfrac{1}{2}\W_{\text{ex}} \vS_1\cdot\vS_2 + \sum_{i=1,2}\W_a (\vS_i\cdot\vec{n}_a)^2$, where $\W_\text{ex}$ is the exchange energy between two sublattices, $\W_a$ is the anisotropy energy, and $\vec{n}_a$ is an anisotropy axis. The non-conservative term can be written as $\mathcal{R}_i = \aG \delta\W/\delta\vS_i - \hbar\sigma j \vec{p}$.

Here we solve the coupled equations~\eqref{eq:LL_coupled} both analytically and numerically. For the analytical solution we formulate a $\sigma$-model, where we substitute the variables $\vS_i$ with N\'{e}el and total spin vectors: $\vl = (\vS_1 - \vS_2)/(S_1 + S_2)$  and $\vs = (\vS_1 + \vS_2)/(S_1 + S_2)$ (see Fig.~\ref{fig:spherical}). We also assume that $|\vs| \ll |\vl|$, which is true in ferrimagnets with antiferromagnetically ordered but unequal sub-lattices near a compensation point, \emph{i.e.}~$S_1/S_2\sim 1$. Also, we assume $\vec{n}_a= \vec{p} = \z$. 

The substitution of the new variables into~\eqref{eq:LL_coupled} produces~\cite{bib:Ivanov:1983}:
\begin{equation}
  \nu\dot{\vl} + \vl\cross\left(\dfrac{1}{\wex}\ddot\vl + \aG\dot\vl + \omega_a\z(\z\cdot\vl) + \tau \z\cross\vl \right) = 0,
  \label{eq:sigma_full}
\end{equation}
where dot stands for a time derivative, $\nu = |(S_1 - S_2)/(S_1 + S_2)| \ll 1$, is the spin-uncompensation constant, $\hbar\wex = \W_\text{ex}$,  $\hbar\omega_a = \W_a$, and $\tau = \sigma j$. Here $\omega_a>0$ corresponds to an easy plane ($\x\y$) and $\omega_a<0$ easy axis ($\z$) anisotropy. The generalized equation~\eqref{eq:sigma_full} transfers to the $\sigma$-model equation for antiferromagnets if $\nu=0$ and to the Landau-Lifshitz equation for ferromagnets, if $S_2\to0$ and $\wex\to\infty$. 

To solve equation~\eqref{eq:sigma_full} we switch to a spherical coordinate system,
$
  \vl = \x \cos\phi\sin\theta + \y \sin\phi\sin\theta + \z\cos\theta
$
(see Fig.~\ref{fig:spherical}). Substituting this decomposition into~\eqref{eq:sigma_full} and projecting on $\z$ and $\x - i\y$ we have:
\begin{gather}
  -\nu \dot \theta \sin\theta +  \frac{\dot \theta \dot \phi}{\wex}\sin2\theta+ \sin^2 \theta \left(\dfrac{\ddot \phi}{\wex}+\aG \dot \phi - \tau\right)=0,\label{eq:sigma_phi}\\
  \nu \dot\phi \sin\theta +
 \frac{\ddot \theta}{\wex} + \aG \dot \theta - \frac{\sin 2 \theta}{2} \left(\frac{\dot \phi^2}{\wex} + \omega_a \right)=0,\label{eq:sigma_theta}
\end{gather}
A steady solution to this system of equations can be found as:
$\phi = \omega_r t$, $\cos\theta = \nu\omega_r\wex/(\omega_r^2 + \omega_a\wex)$.
As one can see from this solution, the N\'{e}el vector $\vl$ rotates around the STT direction, as in the AFM case~\cite{bib:Khymyn:2017,bib:Gomonay:2014,bib:Cheng:2016}, but for $\nu\neq0$ the azimuthal angle $\theta \neq \pi/2$, thus the N\'{e}el vector never fully lies in the $\x,\y$ plane.
We can rewrite the relation between the angle and the rotational frequency in a form:
\begin{equation}
  \cos\theta  =  \tilde\nu \dfrac{\tilde\omega}{\tilde\omega^2 + \sign(\omega_a)},
  \label{eq:theta}
\end{equation}
where $\tilde\omega = \omega_r/\omega_\text{afmr}$, $\omega_\text{afmr} = \sqrt{\wex|\omega_a|}$, and $\tilde\nu = \nu\sqrt{\wex/|\omega_a|}$ is the effective uncompensation parameter. We note here that, $\nu\ll1$, is the small parameter of the model, however, $\tilde\nu$ can be larger than 1.

Equation~\eqref{eq:theta} is the central analytical result of this work: it connects the precession angle $\theta$ with the rotational frequency $\omega_r$. Ultimately, using~\eqref{eq:jsp} and~\eqref{eq:theta} one can compute the output spin-current $\jsp$ as a function of the electric current $j$. The $\x$-component of the spin-current $\jsp$ generates an ac electric field in the Pt layer across the $\z$ direction via the ISHE: $E^\text{AC}_\z = \rho_\perp j e^{-i\omega_r t} = \rho \tSH \jsp\cdot\x e^{-i\omega_r t}$, where $\rho$ is Pt resistivity and $\rho_\perp$ is the Hall-resistivity (see Fig.~\ref{fig:geom}), which can be computed as:
\begin{equation}
  \rho_{\perp}/\rho = \tSH \dfrac{\hbar g_r}{2\pi} \dfrac{\sigma}{\aG}\sin\theta\cos\theta.
  \label{eq:hall}
\end{equation}
This simple expression does not take into account a finite spin-scattering length in Pt and current shunting through the FiM layer~\cite{bib:Althammer:2013, bib:Kim:2016}. Expression~\eqref{eq:hall} can be further simplified if we assume that damping is defined only by spin-pumping, which is typical for thin metallic films ($\aG \approx g_r/(2\pi S_T d_m)$): $\rho_{\perp}/\rho \approx \tSH^2 \sin\theta\cos\theta$.

To illustrate the application of our theory we study a bi-layer of Pt and ferrimagnetic  GdFeCo. For our calculations we use the following parameters~\cite{bib:Zhang:2015, bib:Hirata:2018, bib:Gurevich:1996}: $\tSH = 0.1$, $g_r = \SI{5e18}{m^{-2}}$, $\aG = 10^{-2}$, $S_T = \Ms/(g_e\mu_B) =\SI{5.4e28}{ m^{-3}} $, $\wex/(2\pi) =\SI{3.34}{THz}$,  $\omega_a = \SI{12.6}{GHz}$. The spin-uncompensation parameter in GdFeCo varies with the temperature~\cite{bib:Stanciu:2006, bib:Hirata:2018} and mutual Gd/FeCo concentrations~\cite{bib:Kato:2008,bib:Ostler:2011}. Specially, GdFeCo can be grown in two configurations, easy plane and easy axis, so here we consider two cases: (i) the easy plane is perpendicular to the interface ($\x\y$), and (ii) the easy axis is in the interface plane and along the $\z$-axis. 

\emph{Easy plane}. For a case of an easy plane anisotropy, the ground state of the N\'{e}el vector is $\theta=\pi/2$ and $\phi$ is arbitrary. An application of STT induces an instability, and the N\'{e}el vector starts a rotational motion around the $\z$ axis with the frequency $\omega_r$ defined by~\eqref{eq:omega_r}. Since we do not consider any second anisotropy, this process does not have a current threshold~\cite{bib:Khymyn:2017}. 

\begin{figure}
  % \prntlen{\linewidth}
  \includegraphics{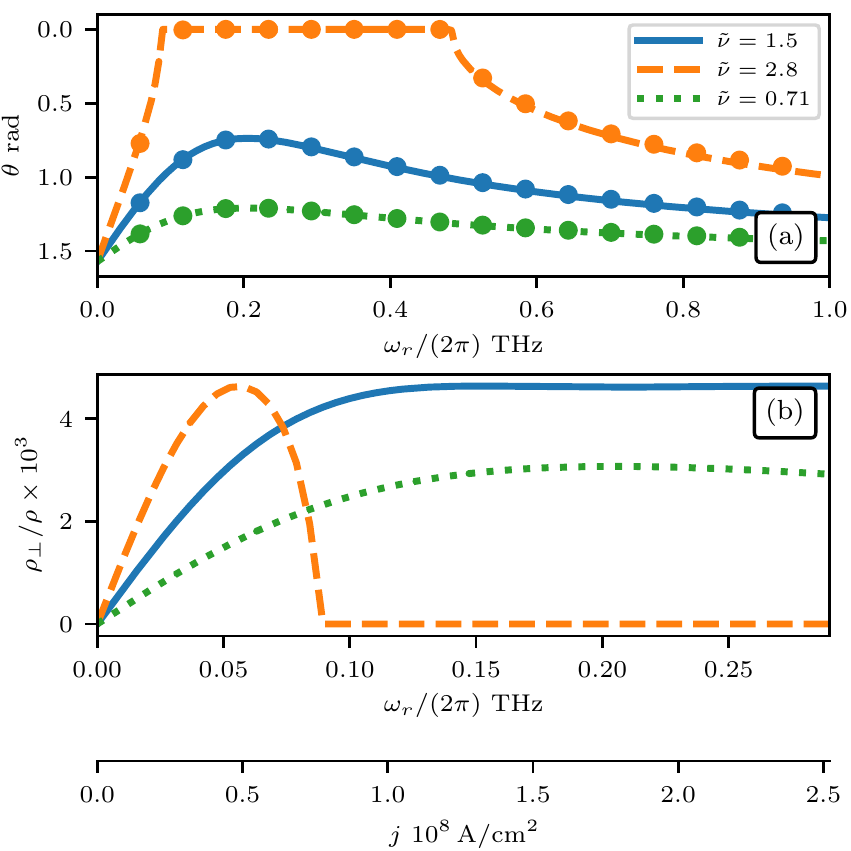}
  \caption{(a) Altitude angle $\theta$ as a function of rotational frequency $\omega_r$ for various values of the effective spin-uncompensation parameter $\tilde\nu$ for the easy plane anisotropy. Lines---analytical solution~\eqref{eq:theta}, dots---numerical solution of~\eqref{eq:LL_coupled}; (b) DC to AC Hall resistance as function of the electric current density in the Pt layer and the rotational frequency $\omega_r$.}\label{fig:theta_ep}
\end{figure}

At the same time, the N\'{e}el vector raises above the $\x\y$ plane. The dependence of the angle of precession (altitude angle) $\theta$ as a function of the rotational frequency, $\omega_r$, is plotted in Fig.~\ref{fig:theta_ep}(a), calculated using formula~\eqref{eq:theta} and by numerically solving~\eqref{eq:LL_coupled}. The analytical and numerical solutions demonstrate practically no discrepancy. For small values of STT, $\sigma j<\aG\omega_\text{afmr}$, the angle of precession approaches zero as the torque increases, since the uncompensated moment tries to align with the torque polarization, and this behavior can be described as FM-like. For large values of torque, $\sigma j>\aG\omega_\text{afmr}$, the precession angle rebounds and tends back to $\pi/2$ as the value of the torque continues to increase. In this regime the AFM-like dynamics prevails. Remarkably, that the small values of $\tilde\nu$ the position of the maximum out-of-plane inclination (minimum $\theta$)  does not depend on $\tilde\nu$: $\omega_r = \omega_\text{afmr}$. The precession angle reaches minimum $\theta_\text{min} = \arccos (\tilde\nu/2)$. 

If the effective uncompensation constant is large, \emph{i.e.}~$\tilde\nu>2$ (note that $\nu$ is still very small), the precession angle can reach the zenith for current densities $j_{1, \text{ep}}^\text{th}<j<j_{2, \text{ep}}^\text{th}$, $j_{i, \text{ep}}^\text{th} = \aG \omega_\text{afmr}/(2\sigma)\left(\tilde\nu \mp \sqrt{\tilde\nu^2 - 4}\right)$. In this situation, all dynamics stops and the N\'{e}el vector fully aligns with $\vec{p}$, similarly to the ferromagnetic case. However, by increasing the values of the STT one can find a point of the AFM-like instability $j^\text{th}_{2, \text{ep}}$ after which the dynamics resumes. 

The induced Hall resistivity is plotted in~Fig.~\ref{fig:theta_ep}(b). Formula~\eqref{eq:hall} suggests that the maximum output voltage occurs for $\theta=\pi/4$, thus it is derisible to have $\tilde\nu \approx \sqrt{2}$. However, this exact value can be challenging to achieve experimentally. Fortunately, our calculations show, that the value of the Hall resistivity is quite robust against varying $\tilde\nu$: decrease of $\tilde\nu$ from the ``ideal value'' leads of a gradual reduction of $\rho_\perp$, although, increase of $\tilde\nu$ leads to reducing the bandwidth where the Hall resistance is maximized.

\emph{Easy axis}. In the easy axis configuration the N\'{e}el vector has two ground states $\theta = 0,\pi$. In a compensated AFM case ($\nu=0$) these states are fully equivalent. However, the spin uncompensation removes this degeneracy with respect to inversion along the $\z$-axis, which leads to a different behavior of the N\'{e}el vector for opposite directions of the STT. In contrast with the compensated AFM and fully uncompensated FM cases, here two types of instabilities may exist with respect to the STT strength. One type of instability leads to a FM-like $\pi$ radian revolution (switching) of the N\'{e}el vector from one static equilibrium to another. Another type of instability leads to a steady AFM-like rotation of the N\'{e}el vector.  

\begin{figure}
  \includegraphics{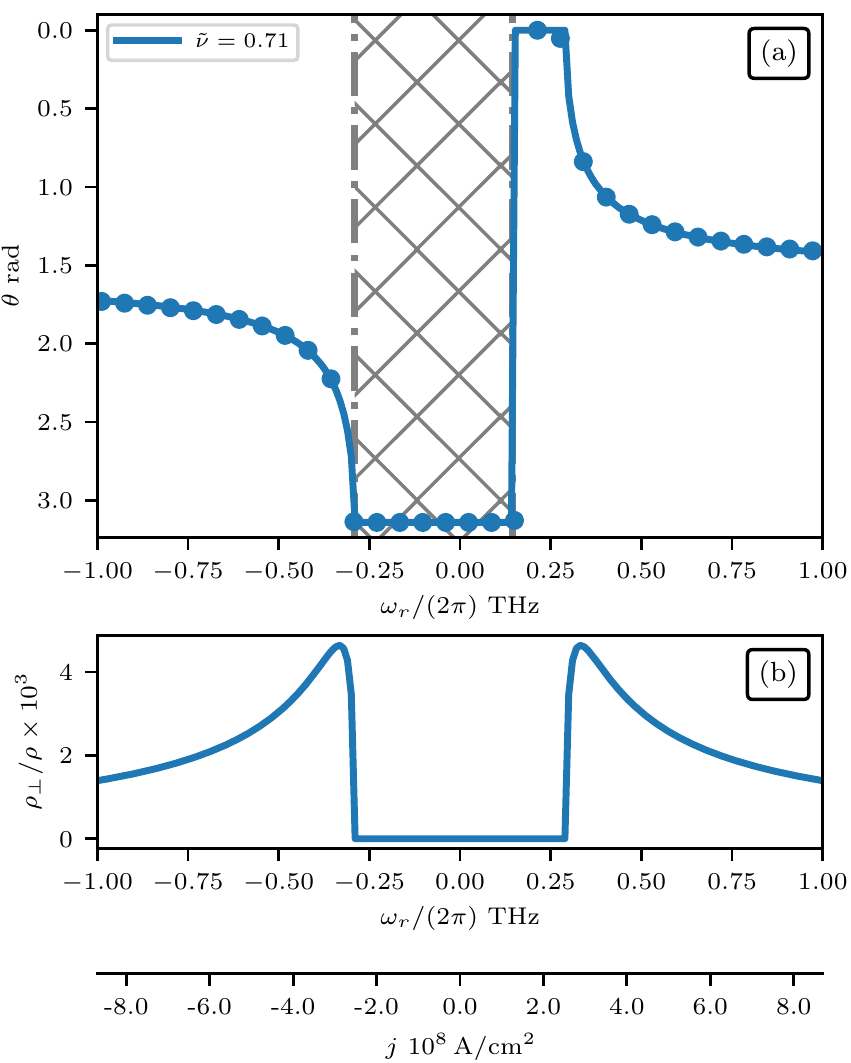}
  \caption{(a) Equilibrium values of the altitude angle $\theta$ as a function of the rotational frequency $\omega_r$ for an initial state $l_z = -1$ and the easy axis anisotropy. Lines---analytical solution, dots---numerical solution of~\eqref{eq:LL_coupled}. Hatching denotes the region of current densities for which ground state $l_z=-1$ is stable, calculated from~\eqref{eq:ea_th}. To get the same diagram for $l_z=1$ one should apply $\omega_r \to-\omega_r$ and $\theta \to \pi - \theta$ (b) DC to AC Hall resistance as function of the electric current density in the Pt layer and the rotational frequency $\omega_r$.}\label{fig:theta_ea}
\end{figure}

To find the thresholds of these instabilities we analyze the stability of~\eqref{eq:sigma_full} near the ground state $\vl_0 = \pm\z$. Near the equilibrium point we can decompose the N\'{e}el vector as: $\vl(t) = l_z\z + l_x(t)\x+l_y(t)\y$, where $l_z = \pm 1$. Substituting this expression into~\eqref{eq:sigma_full} and using $l_x(t) + i l_y(t) = \Psi_0 e^{\Gamma t}$ one obtains:
\begin{equation}
  i \nu \Gamma l_z = \dfrac{\Gamma^2}{\wex}  + \wa + \aG \Gamma - i \tau
\end{equation}
Here we see that an instability develops when $\Real{\Gamma} > 0$. Solving the former equation and using~\eqref{eq:omega_r} we find the threshold current density sufficent for an instability:
\begin{gather}
j^\text{th}_\pm = \dfrac{\aG\wafmr}{2\sigma}\left(\pm \sqrt{4+\tilde\nu^2} +l_z \tilde\nu\right)
\label{eq:ea_th}
\end{gather}
These two thresholds correspond to two processes: (i) an AFM-like instability, when the N\'{e}el vector comes to a steady motion~\cite{bib:Gomonay:2014, bib:Cheng:2016, bib:Khymyn:2017}, and (ii) an FM-like anti-damping switching of the N\'{e}el vector from one equilibrium to another~\cite{bib:Chen:2018}. 

The dependence of the azimuthal angle for different regimes for the ground state $l_z=-1$ is illustrated in~Fig.~\ref{fig:theta_ea}(a). Under a large negative current density $j<j^\text{th}_-$ the N\'{e}el vector experiences an infinite conical precession with angle $\theta$ defined by~\eqref{eq:theta}. In contrast, if the current density is $j^\text{th}_+<j<-j^\text{th}_-$ the N\'{e}el vector switches its ground state orientation from $l_z=-1$ to $l_z=1$. Further, if the current density overcomes $j>-j^\text{th}_-$ the N\'{e}el vector start a precessional motion.  

\begin{figure}
  \includegraphics{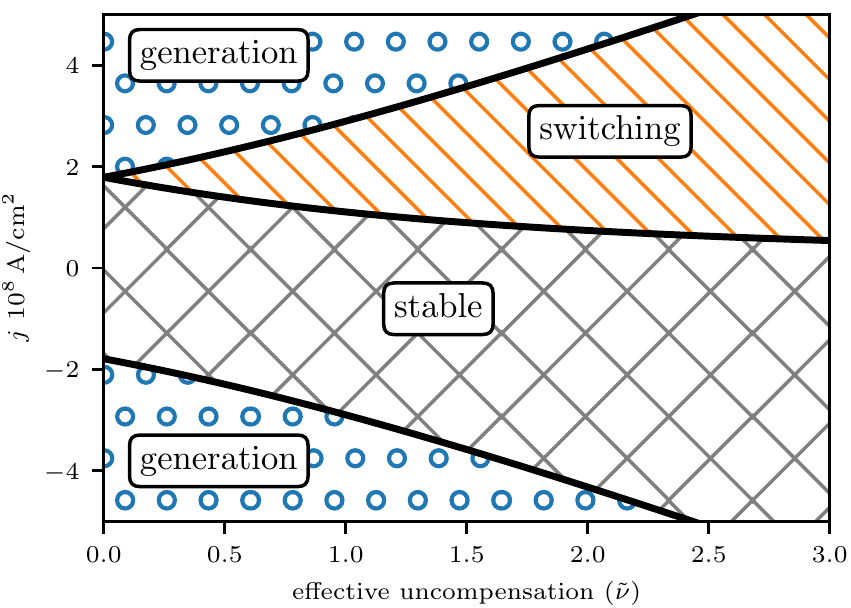}
  \caption{Phase diagram of instabilities in easy axis ferrimagnet under spin-transfer torques as a function of the spin uncompensation and the current density for $l_z=-1$ ground state.}
  \label{fig:phase_diagram}
\end{figure}

The values of threshold current densities strongly depend on the uncompensation constant. The ``phase diagram'' of the switching and generation instabilities is shown in Fig.~\ref{fig:phase_diagram}. The spin uncompensation increases the ``generation threshold'' $j^\text{th}_-$  and decreases the ``switching threshold'' $j^\text{th}_+$. Therefore, for the generation regime the FiM should be mostly compensated. However, a small uncompensation is necessary for the conical precession of the N\'{e}el vector for a non-zero spin-Hall resistance (see Fig.~\ref{fig:theta_ea}(b)). Thus, for the easy axis anisotropy, one should choose a compromise between the generation threshold current and the bandwidth of the generation. We also note that for an ideally compensated AFM ($\nu=0$)  $|j^\text{th}_\pm| = \aG\sigma^{-1}\omega_\text{afmr}$ found in~\cite{bib:Gomonay:2010, bib:Cheng:2016} for an easy axis AFM.

Although we considered a particular ``spin-Hall'' geometry of a Pt/GdFeCo bi-layer, our theory can be applied to standard nano-pillar~\cite{bib:Kim:2012} and nano-contact~\cite{dumas2014recent} STNOs where the ferromagnetic layer is substituted by a ferrimagnetic layer. Since the electronic structure of rare-earth and transition metals is substantially different, we expect a non-zero magnetoresistance in GdFeCo even at the spin-compensation point. In the STNO geometry the tensor of magnetoresistance has a similar form as~\eqref{eq:jsp}, albeit with a different coefficient, thus the conical precession will allow for a non-zero AC magnetoresitance component.

In conclusion, we developed a theory of ferrimagnetic dynamics under a spin-transfer torque. We demonstrated that the precession of the N\'{e}el vector in ferrimagnets is conical and, in contrast with the antiferromagnetic case, this precession generates a non-zero AC spin current. We have shown that in a case of small non-compensation it is possible to achieve sub-THz frequency precession in typical GdFeCo ferrimagnetic allosy in both cases of easy plane and easy axis anisotropy.  

\begin{acknowledgments}
BAI gratefully acknowledges support from the National Academy of Sciences of Ukraine via Project No. 1/17 H, the Program of NUST ``MISiS'' (Grant No. K2--2019--006), implemented by the Russian Federation governmental decree dated 16th of March 2013, No. 211; and by the department of Targeted Training of Taras Shevchenko National University of Kyiv at the National Academy of Sciences of Ukraine via project ``Elements of ultrafast neuron systems on the basis of antiferromagnetic spintronic nanostructures''. RK and J\r{A} acknowledge support from the Knut and Alice Wallenberg Foundation.
\end{acknowledgments}

\bibliography{FiM.bib}

\end{document}